\documentclass[aip,apl,amsmath,amssymb,reprint]{revtex4-1}
\usepackage{graphicx}
\usepackage{dcolumn}

\usepackage[utf8]{inputenc}
\usepackage[T1]{fontenc}
\usepackage{mathptmx}

\begin{document}

\title{Anisotropic ultrafast optical response of terahertz pumped graphene}

\author{A. A. Melnikov}
\email{melnikov@isan.troitsk.ru}
\affiliation{Institute for Spectroscopy, Russian Academy of Sciences, 108840 Troitsk, Moscow, Russia}
\affiliation{National Research University Higher School of Economics, 101000 Moscow, Russia}

\author{A. A. Sokolik}
\affiliation{Institute for Spectroscopy, Russian Academy of Sciences, 108840 Troitsk, Moscow, Russia}
\affiliation{National Research University Higher School of Economics, 101000 Moscow, Russia}

\author{A. V. Frolov}
\affiliation{Kotelnikov Institute of Radioengineering and Electronics, Russian Academy of Sciences, 125009 Moscow,
Russia}

\author{S. V. Chekalin}
\affiliation{Institute for Spectroscopy, Russian Academy of Sciences, 108840 Troitsk, Moscow, Russia}

\author{E. A. Ryabov}
\affiliation{Institute for Spectroscopy, Russian Academy of Sciences, 108840 Troitsk, Moscow, Russia}

\begin{abstract}
We have measured the ultrafast anisotropic optical response of highly doped graphene to an intense single cycle
terahertz pulse. The time profile of the terahertz-induced anisotropy signal at 800 nm has minima and maxima repeating those of the pump terahertz
electric field modulus. It grows with increasing carrier density and demonstrates a specific nonlinear dependence on
the electric field strength. To describe the signal, we have developed a theoretical model that is based on the energy
and momentum balance equations and takes into account optical phonons of graphene and substrate. According to the
theory, the anisotropic response is caused by the displacement of the electronic momentum distribution from zero
momentum induced by the pump electric field in combination with polarization dependence of the matrix elements of
interband optical transitions.
\end{abstract}

\maketitle

Due to the peculiar electronic band structure of graphene \cite{CastroNeto,DasSarma} the field-induced motion of
electrons was predicted to be strongly nonlinear in this material \cite{Mikhailov}. High nonlinearity together with the
unique electronic and optical properties make graphene a prospective material for photonic and optoelectronic
applications. In the light of this perspective, nonlinear optical phenomena in graphene are actively studied
\cite{Bao,Bonaccorso, Ooi}. Among them are harmonic
generation\cite{Hong,Cheng,Kumar,Soavi,Jiang,Taucer,Baudisch,Yoshikawa,Bowlan,Hafez2018}, saturable absorption
\cite{Winzer,Marini,Cihan}, self-phase modulation \cite{Vermeulen}, and four-wave mixing \cite{Hendry,Koenig}. In the optical
range the frequency of light is higher than the electron-electron scattering rate, so the resulting ``coherent''
electronic response is determined by the properties of single-electron band structure \cite{Taucer,Baudisch,Yoshikawa}.
In the THz range another limiting case is realized --- the characteristic time of electron-electron scattering
processes is shorter than the period of the light wave. The energy imparted by the electric field to electrons is
quickly redistributed heating the electron gas, while the electron-phonon collisions cool and decelerate the gas. The
concept of ``incoherent'' nonlinearity that appears due to the change of electron gas conductivity upon heating
\cite{Mics,Razavipour} was employed recently to explain highly effective generation of THz harmonics in graphene
\cite{Hafez2018}.

The routine technique used to evaluate the optical nonlinearity of graphene is the spectral analysis of light
transmitted through the sample in search of harmonics of the pump frequency radiated by the nonlinear current. In the
present work we employ an alternative approach by using an optical probe to detect the transient THz field-induced
shift of electron momentum distribution (note that such shift can induce the optical 2-nd harmonic generation, as was
recently observed \cite{Tokman}). In graphene, due to the specific polarization dependence of matrix elements of
interband transitions in combination with Pauli blocking, an anisotropy of electronic distribution implies an
anisotropy of infrared optical conductivity, which can be measured by detecting depolarization of probe light reflected
from the sample. We measure the ultrafast anisotropic optical response of graphene to intense THz pulses and show that
though the corresponding signal is rather weak, it can be reliably detected for heavily doped graphene and contains
specific nonlinear features. To interpret the signal, we develop a model based on the Boltzmann kinetic equation solved
in the hydrodynamic approximation.

The sample used in our experiments was a sheet of single-layer CVD graphene on the SiO$_2$/Si substrate
(the thickness of SiO$_2$ was 300 nm). Four indium contacts were attached to the sample in order to apply gate voltage and to measure the resistance of the graphene layer. Nearly single-cycle THz pulses with a duration of about 1 ps were generated in a
lithium niobate crystal in the process of optical rectification of femtosecond laser pulses with tilted fronts (see,
e.g., Ref. \cite{Stepanov} for details). The THz generation stage was fed by 50~fs laser pulses at 800 nm, 1.2 mJ per
pulse at 1 kHz repetition rate. THz radiation was focused by a parabolic mirror so that the peak electric field of the
THz pulses incident on the sample was $\sim$ 400 kV/cm (denoted below as $E_\mathrm{max}$). The waveform of the pulses
was characterized by means of electro-optic detection in a 0.15 mm thick (110)-cut ZnTe crystal. The central frequency of the THz pulse was $\sim$ 1.5 THz, while its spectral width $\sim$ 2 THz (FWHM). In the experiments we detected transient anisotropic changes of reflectance of the sample caused by the pump THz pulses. The probe 50~fs
pulses at 800 nm were polarized before the sample at 45$^{\circ}$ relative to the vertical polarization of pump THz
pulses. Both pump and probe beams were incident onto the sample at an angle of $\sim$ 7$^{\circ}$. Upon reflection from
the sample excited by THz radiation the probe pulses experienced a small rotation of polarization, which was detected
by measuring the intensities of two orthogonal polarization components of the reflected probe beam $I_{r,x}$ and
$I_{r,y}$ using a Wollaston prism and a pair of photodiodes. The quantity
\begin{eqnarray}
F=1-\frac{I_{r,y}}{I_{r,x}}.\label{F}
\end{eqnarray}
as a function of the probe pulse delay time is referred to as anisotropy signal or anisotropic response.

\begin{figure}
\begin{center}
\resizebox{1.0\columnwidth}{!}{\includegraphics{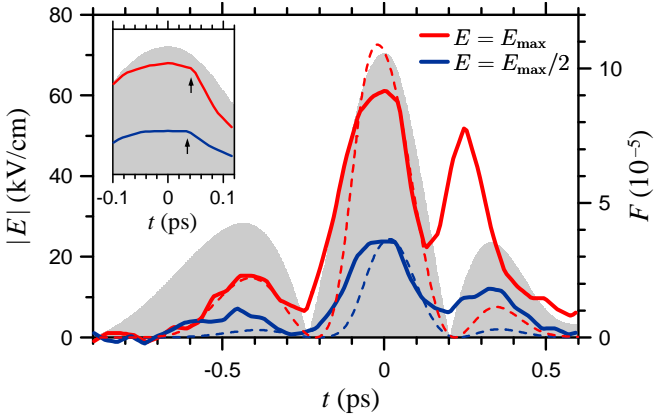}}
\end{center}
\caption{\label{Fig1}Optical anisotropy signals $F(t)$ measured at the peak THz fields $E_\mathrm{max}$ (top solid
line) and $E_\mathrm{max}/2$ (bottom solid line). The shaded area depicts temporal profile of the field modulus
$|E(t)|$. Dashed lines show calculation results at the peak THz field strength of $70\,\mbox{kV/cm}$ (top dashed line) and
$35\,\mbox{kV/cm}$ (bottom dashed line). The inset shows a magnified view of the central peaks of $F(t)$ and of the THz
field (the arrows point at kinks in the $F(t)$ waveforms).}
\end{figure}

The anisotropic response of the sample induced by the pump THz pulse is shown in Fig.~\ref{Fig1} for the peak values of
the electric field of $E_\mathrm{max}$ and $E_\mathrm{max}/2$. As soon as the signal from the regions of SiO$_2$/Si
substrate not covered by graphene was below the noise level, we concluded that the source of the observed anisotropy
signal was the graphene layer itself. Fig.~\ref{Fig1} also shows the temporal profile of the electric field of the pump
THz pulse. In order to ensure linearity of the electrooptic detection in the ZnTe crystal we attenuated pump THz beam
power by a factor of $\sim$ 400 using a variable metallic filter. To record the electric field profile the filter was
``closed'' so that the THz beam passed through the fused silica plate covered by the thickest metallic layer. The
sample response at $E_\mathrm{max}/2$ was measured with the ``opened'' filter as the pump THz pulses passed only
through fused silica (the 2 mm thick fused silica plate reduces the THz field by a factor of $\approx2$). Finally, in
order to detect the anisotropic response induced by the strongest pump electric field available ($E_\mathrm{max}$) we
removed the variable filter so that the THz radiation traveled to the sample only through air. As soon as the fused
silica plate causes a large additional retardation of THz pulses the anisotropic response measured at $E_\mathrm{max}$
was time-shifted so that it matched the signal detected at $E_\mathrm{max}/2$ in time domain. As follows from
Fig.~\ref{Fig1} the third peak in the signal measured at $E_\mathrm{max}$ occurs earlier than the corresponding peaks
in the signal detected at $E_\mathrm{max}/2$ and in the electric field profile $|E(t)|$. This effect is due to group velocity
dispersion in the fused silica plate that leads to a $\sim$ 20\% lengthening of the pulse and of the signal. 
Variation of the relative amplitude of the third peak in $|E(t)|$ caused by the plate is negligible.

\begin{figure}
\begin{center}
\resizebox{1.0\columnwidth}{!}{\includegraphics{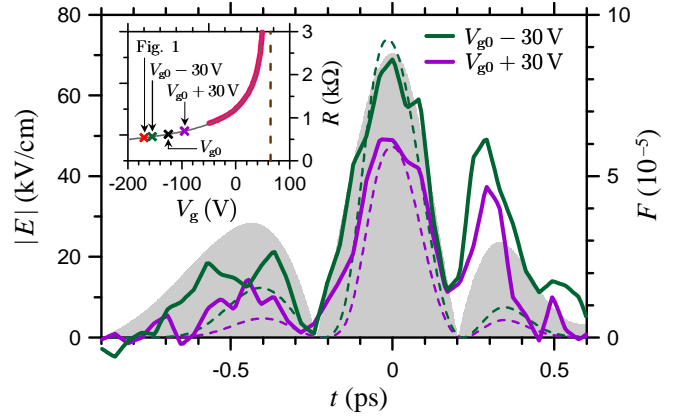}}
\end{center}
\caption{\label{Fig2}Optical anisotropy signals $F(t)$ measured at full THz field strength for two values of the total
effective gate voltage: $V_{\mathrm{g}0}-30\,\mbox{V}$ (top solid line) and $V_{\mathrm{g}0}+30\,\mbox{V}$ (bottom
solid line). Dashed lines show calculation results at the peak THz field strength of $70\,\mbox{kV/cm}$ for the higher
carrier density (top dashed line, $\mu=480\,\mbox{meV}$) and for the lower density (bottom dashed line,
$\mu=430\,\mbox{meV}$). The inset shows the measured resistance of the graphene sample as a function of $V_\mathrm{g}$
(thick line) and its analytical approximation (thin line) with the charge neutrality point location (dashed line), and
with the estimated doping levels during the pump-probe experiments shown by crosses.}
\end{figure}

To estimate the doping level of graphene, we measured the resistance of the sample as a function of gate voltage
$V_\mathrm{g}$ soon after its preparation (thick line in the inset to Fig.~\ref{Fig2}). We approximated this dependence by the formula
$R(V_\mathrm{g})\approx
R_0+A/|V_\mathrm{g}-V_\mathrm{CNP}|+B/|V_\mathrm{g}-V_\mathrm{CNP}|^{3/2}+C/|V_\mathrm{g}-V_\mathrm{CNP}|^2$ with the
charge neutrality point location $V_\mathrm{CNP}\approx65\,\mbox{V}$, which takes into account short- and long-range
impurities \cite{DasSarma} (the first two terms), and corrections proportional to higher powers of the inverse Fermi
momentum (the last two terms). This approximation was extrapolated to the measured values of $R$, allowing us to
estimate the current Fermi level position $\mu$. We found that when the ultrafast measurements were performed several months later, 
the Fermi level of graphene shifted considerably probably due to doping by water molecules adsorbed from ambient air. 
The shift was of such magnitude as if the effective gate voltage $V_{\mathrm{g}0}\approx-125\,\mbox{V}$ was applied. 
Application of the real gate voltages $\mp30\,\mbox{V}$, which
were effectively added to $V_{\mathrm{g}0}$ resulting in the total effective gate voltages
$V_{\mathrm{g}0}\mp30\,\mbox{V}$, allowed us to increase (decrease) the charge carrier concentration, leading to
increase (decrease) of the anisotropy signal. The experiment illustrated by Fig.~\ref{Fig1} was performed even later than the one,
 the results of which are shown in Fig.~\ref{Fig2}. The doping level is this case was estimated as $\mu\approx-500\,\mbox{meV}$,
corresponding to the hole density $n\approx2\times10^{13}\,\mbox{cm}^{-2}$. (In calculations below we assume positive
$\mu$ for better clarity, because our model is particle-hole symmetric).

Time evolution of the electron gas in highly doped graphene under intense THz field $\mathbf{E}(t)$ is dominated by its
intraband dynamics \cite{Tani,Hafez,Tomadin}, described in terms of two separate momentum distribution functions
$f_\pm(\mathbf{k},t)$ for electrons in conduction and valence bands. Time evolution of these functions is described by
the semiclassical Boltzmann kinetic equation
\begin{eqnarray}
\frac{\partial f_\gamma}{\partial t}=-\frac{e\mathbf{E}}\hbar\cdot\frac{\partial f_\gamma}{\partial\mathbf{k}}
+\frac{\langle f_\gamma\rangle+\mathbf{n}\cdot\langle\mathbf{n}f_\gamma\rangle-f_\gamma}{\tau_\mathrm{imp}(k)}
\nonumber\\+\Gamma_\gamma^\mathrm{in}(1-f_\gamma)-\Gamma_\gamma^\mathrm{out}f_\gamma+ \left(\frac{\partial
f_\gamma}{\partial t}\right)_\mathrm{ee}.\label{Boltzmann}
\end{eqnarray}
The terms in the right hand side describe, respectively, electron acceleration by the applied electric field, elastic
collisions with impurities \cite{Kashuba} with momentum-dependent scattering time $\tau_\mathrm{imp}(k)$,
electron-phonon and electron-electron collisions. $\Gamma_\gamma^\mathrm{in,out}(\mathbf{k},t)$ are the rates of electron scattering
into the $\mathbf{k}\gamma$ state and out of this state \cite{Malic,Kim,Brida}.

\begin{figure}[t]
\begin{center}
\resizebox{1.0\columnwidth}{!}{\includegraphics{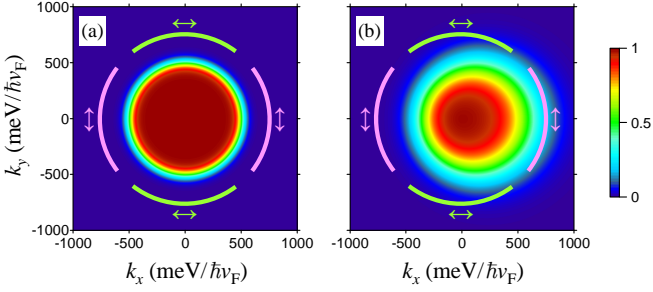}}
\end{center}
\caption{\label{Fig3}Momentum distribution functions $f_+(\mathbf{k})$ of electrons in the conduction band of graphene
in the hydrodynamic model (\ref{drift}). (a) The initial state of the electron gas with $T=300\,\mbox{K}$,
$\mu=500\,\mbox{meV}$, $V_x=0$, (b) The moving heated gas with $T=1000\,\mbox{K}$, $\mu=450\,\mbox{meV}$
(reduced to ensure particle number conservation), $V_x=0.24v_\mathrm{F}$. The arcs schematically depict the electron
states, preferentially involved in the interband optical transitions at $\hbar\omega_\mathrm{pr}=1.55\,\mbox{eV}$ with
the linear polarizations shown by the arrows.}
\end{figure}

Interband dynamics of the electron gas induced by the THz field in our case should be slow with respect to fast
electron-electron collisions, which thermalize the electron gas on a time scale less than $30\,\mbox{fs}$
\cite{Johanssen,Gierz}. Consequently, $f_\gamma(\mathbf{k},t)$ can be taken in the form of the ``hydrodynamic''
distribution function
\begin{eqnarray}
f^\mathrm{drift}_\gamma(\mathbf{k},t)=\left\{\exp\left[\frac{\epsilon_{k\gamma}-\hbar\mathbf{k}\cdot\mathbf{V}(t)
-\mu(t)}{T(t)}\right]+1\right\}^{-1},\label{drift}
\end{eqnarray}
which is formed due to electron-electron collisions with conservation of total energy, momentum and particle number
\cite{Kashuba,Briskot}. Here $\epsilon_{k\gamma}=\gamma v_\mathrm{F}k$ are the single-particle energies, while
temperature $T$, chemical potential $\mu$, and drift velocity $\mathbf{V}$ are slowly varying functions of time.
Fig.~\ref{Fig3} shows examples of (\ref{drift}) for $n$-doped graphene. Note that owing to the linear dispersion in
graphene the distribution function (\ref{drift}) is not just a shifted Fermi sphere, as it would be in the case of massive electrons, but rather a
gas with anisotropic temperature. Combined action of the strong THz field that accelerates electrons and rapid thermalization 
makes the distribution function elongated in the direction of the THz field, while the subsequent impurity and phonon
scattering tends to make $f_\gamma(\mathbf{k},t)$ isotropic, leading to electron gas heating.

A nonzero drift velocity $V_x$ (we take $\mathbf{E}$ and $\mathbf{V}$ along the $x$ axis) makes the distribution
functions (\ref{drift}) angular anisotropic at the probe pulse wave vector modulus
$|\mathbf{k}|=k_\mathrm{pr}=\omega_\mathrm{pr}/v_\mathrm{F}$. In combination with the angular dependence of the matrix
elements of interband transitions \cite{Malic}, it leads to the anisotropy of the optical conductivity tensor at
$\omega=\omega_\mathrm{pr}$:
\begin{eqnarray}
\left\{\begin{array}{ll}\sigma_{xx}\\
\sigma_{yy}\end{array}\right\}=\frac{e^2}{4\pi\hbar}\int\limits_0^{2\pi}d\varphi\:
\left\{\begin{array}{lr}\sin^2\varphi\\
\cos^2\varphi\end{array}\right\}\left.(f_--f_+)\right|_{k=k_\mathrm{pr}}.
\end{eqnarray}

The small difference between $\sigma_{xx}$ and $\sigma_{yy}$ manifests itself in the reflectances $R_{x,y}$ of the
whole graphene/SiO$_2$/Si structure for the $x$- and $y$-polarized probe pulses at normal incidence. Defining the
optical contrast of graphene on a substrate as $C\approx-(\sigma/R)(\partial R/\partial\sigma)$
\cite{Blake,Casiraghi,Fei}, we can calculate the anisotropy signal (\ref{F}) as
\begin{eqnarray}
F=\frac{R_x-R_y}{R_x}\approx C\frac{\sigma_{yy}-\sigma_{xx}}{e^2/4\hbar}\nonumber\\
=\frac{C}{\pi}\int\limits_0^{2\pi}d\varphi\:\cos2\varphi\left.(f_--f_+)\right|_{k=k_\mathrm{pr}}.\label{F_theor}
\end{eqnarray}
For graphene on Si covered by the $300\,\mbox{nm}$-thick $\mathrm{SiO}_2$ layer, the optical contrast at
$\lambda_\mathrm{pr}=800\,\mbox{nm}$ is rather small and negative \cite{Casiraghi}. Calculating it using the transfer
matrix method \cite{Fei}, which allows us to take into account multiple reflections from graphene and Si substrate, and
taking the universal optical conductivity of graphene $\sigma=e^2/4\hbar$ in the calculation, we get $C\approx-0.0044$.
In principle, by adjusting the $\mathrm{SiO}_2$ layer thickness \cite{Blake} in order to enhance the visibility of
graphene it is possible to increase the observed signal.

The physical origin of the optical anisotropy is illustrated by Fig.~\ref{Fig3}(b). The interband transitions for the
$y$-polarized light become suppressed with respect to those for the $x$ polarization due to Pauli blocking, caused by
the thermal tail of the displaced distribution function at $V_x\neq0$. The resulting difference of the conductivities,
$\sigma_{yy}<\sigma_{xx}$, leads to a positive anisotropy signal (\ref{F_theor}) since $C$ is negative. This picture is
symmetric when $V_x$ changes sign, so in the limit of low drift velocity $F\propto V_x^2$. Unlike studies with linearly
polarized optical pump \cite{Mittendorff,Trushin,Malic,Yan1,Danz,Konig-Otto}, where momentum distribution of the
photoexcited electrons and holes is highly anisotropic ($\sim\sin^2\varphi$) from the very beginning in spite of the
zero total momentum, in our case the anisotropy arises as the electron Fermi sphere is displaced from zero momentum by
the strong THz field. In both cases the distribution functions acquire nonzero second angular harmonics
($\sim\cos2\varphi$) that is necessary for the anisotropy of the optical response.

We solve the Boltzmann equation (\ref{Boltzmann}) in the hydrodynamic approximation (\ref{drift}), using balance
equations for the total energy, momentum and particle number of the electron gas similarly to the works on electron
transport in graphene in stationary high electric fields \cite{Bistritzer,DaSilva}. In these equations, the energy and
momentum time derivatives caused by phonons are calculated using full electron-phonon collision integrals. We consider
6 phonon modes: 4 modes of graphene $\boldsymbol\Gamma$ and $\mathbf{K}$ optical phonons
\cite{Malic,Kim,Brida,Butscher} and 2 modes of $\mathrm{SiO}_2$ surface polar phonons
\cite{Fratini,Konar,Perebeinos,Yan2}. We assume polarization- and momentum-independent phonon occupation numbers
$n_\mu=[\exp(\hbar\omega_\mu/T_\mu)-1]^{-1}$ determined by two separate temperatures for graphene optical
$T_\mathrm{GO}$ and surface polar $T_\mathrm{SPP}$ phonons. Since hot phonons play an important role in the electron
gas dynamics in strong fields \cite{Malic,Butscher,Perebeinos}, we calculate time evolution of $T_\mathrm{GO}$ and
$T_\mathrm{SPP}$ from the energy balance for the corresponding phonon gases, which exchange energy with electrons and additionally
lose energy via phonon decay with the characteristic times $\tau_\mathrm{ph}\approx 2\,\mbox{ps}$
\cite{Johanssen,Gierz} and $\tau_\mathrm{SPP}\approx1\,\mbox{ps}$ \cite{Yan2}. For the scattering time on long-range
impurities, relevant for graphene on a $\mathrm{SiO}_2$ substrate \cite{Al-Naib}, we take $\tau_\mathrm{imp}(k)\approx
s/k$, where $s$ can be related to the low-field carrier mobility $\mu_\mathrm{c}=2ev_\mathrm{F}s/\hbar$, which is about
$1000\,\mbox{cm}^2/\mbox{V}\cdot\mbox{s}$ in our sample. In numerical calculations we use the THz electric field
strength of $\sim70\,\mbox{kV/cm}$ that is several times lower than the incident field $E_\mathrm{max}$. This reduction
of the field acting on graphene electrons is caused by the destructive interference of the incident THz wave with the
wave reflected from the underlying $p$-doped Si substrate with the reflectivity $R_\mathrm{THz}\approx0.6\div0.7$
\cite{Ray}.

Typical calculation results for $F$ are shown in Fig.~\ref{Fig1}. We take the doping level $\mu=500\,\mbox{meV}$ and
the values $70\,\mbox{kV/cm}$ and $35\,\mbox{kV/cm}$ for the peak strength of the electric field acting on graphene
electrons. One can see that our numerical model reproduces the magnitude of $F$ and the general similarity of $F(t)$
and $|E(t)|$ relatively well at realistic parameters. Fig.~\ref{Fig2} illustrates the dependence of the calculated $F$
 on the doping level. For the calculation we used Fermi levels $\mu=430$ and $480\,\mbox{meV}$ for the cases of lower and higher doping. 
 These values were close to those extracted from the resistance measurements and allowed us to reproduce the experimental results relatively well. 
 Both the theory and the experiment demonstrate the same qualitative effect --- $F$ grows with increasing doping level. However, generally the theory predicts 
 highly nonlinear doping dependence of $F$, especially for strong THz fields. Note that
increasing $|\mu|$ or decreasing $\omega_\mathrm{pr}$ in order to bring optically probed energy regions
$\pm\hbar\omega_\mathrm{pr}/2$ closer to the Fermi level would significantly increase the anisotropy signal.

\begin{figure}
\begin{center}
\resizebox{1.0\columnwidth}{!}{\includegraphics{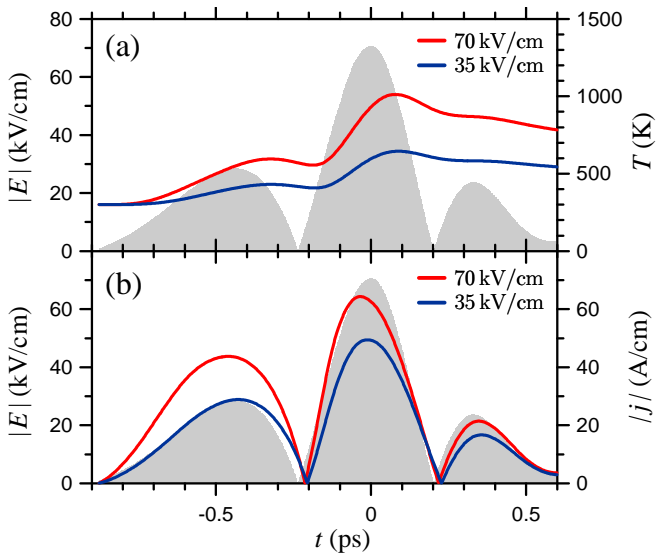}}
\end{center}
\caption{\label{Fig4}Calculated electron gas temperatures (a) and electric current densities (b) at the peak THz field strength
of $70\,\mbox{kV/cm}$ (top curves) and $35\,\mbox{kV/cm}$ (bottom curves).}
\end{figure}

The detected anisotropic response of graphene, however, contains specific features, that our model is not able to
reproduce. First, the third peak behaves differently with respect to the first two ones: its growth upon doubling the
electric field is considerably higher ($\sim4.5$) than for the first two peaks ($\sim2.5$), and is underestimated by
the theory. This anomalous behavior at the end of the THz pulse can be caused by the heating of the electron gas in
graphene, the temperature of which is expected to be maximal after the action of the peak electric field, as shown by
the calculated profiles of $T(t)$ in Fig.~\ref{Fig4}(a). It should also be noted that the rise time of the third peak
is the shortest of all three peaks ($\sim50\,\mbox{fs}$) and is comparable with the characteristic time of
electron-electron interactions in graphene, so in this regime our hydrodynamic approximation (\ref{drift}) can miss
some features of the coherent collisionless dynamics of electrons driven by the high electric field.

One more interesting property of the anisotropic response is the sharp bend or kink observed in the signal at
$\sim40\,\mbox{fs}$, after the THz electric field has reached the maximum and just began to decrease. It is visible in
the signals recorded at both field strengths and is marked by arrows in the inset to Fig.~\ref{Fig1}. One can see that
due to this kink the form of the anisotropy signal differs considerably from the THz waveform. The latter evolves
smoothly similar to a sine wave, while the former resembles a wave crest indicating the nonlinearity of the THz
response of graphene. Such behavior of the anisotropic signal near the peak electric field can be a signature of
similar nonlinear features in the THz-induced current, although we do not measure the latter directly in our experiment.

Finally, in view of the long-standing search of the nonlinear current response of graphene in the THz range
\cite{Mikhailov,Bowlan,Hafez2018}, we calculate the electric current density $j$ (Fig.~\ref{Fig4}(b)). The electric
current demonstrates strong nonlinearities: first, its peak values change insignificantly when the electric field is
doubled, that can be considered as a manifestation of the electric current saturation \cite{DaSilva,Perebeinos}, and,
second, $j$ becomes lower at the same field strength near the end of the pulse, which can be attributed to the
influence of electron gas heating.

In conclusion, we have measured the ultrafast anisotropic optical response of highly doped graphene under intense THz
excitation and developed the model of temporal dynamics of the momentum distribution functions based on the Boltzmann
equation, solved in the hydrodynamic approximation. Theoretical calculations provide good description of the general
shape and magnitude of the anisotropy signal at realistic parameters, and also predict strong nonlinearities of the
THz-field induced electric current. We demonstrate that the anisotropic optical response measured with subcycle
temporal resolution contains information on the ultrafast dynamics of the electron gas, its heating, isotropization and
concomitant nonlinearities. Our work links the areas of nonlinear THz electrodynamics of graphene \cite{Mikhailov},
ultrafast pseudospin dynamics of Dirac electrons \cite{Mittendorff,Trushin,Malic,Yan1,Danz,Konig-Otto}, and
strong-current graphene physics \cite{DaSilva,Perebeinos}, thereby providing an alternative tool for studying high-field
phenomena in graphene in the far-IR and THz range.

This work was supported by the Ministry of Education and Science of the Russian Federation (Project No.
RFMEFI61316X0054). The experiments were performed using the Unique Scientific Facility ``Multipurpose femtosecond
spectroscopic complex'' of the Institute for Spectroscopy of the Russian Academy of Sciences.

\end{document}